\documentclass[superscriptaddress,twocolumn,amsmath,amssymb]{revtex4-1}

\usepackage{epsf}
\usepackage{epsfig}
\usepackage{latexsym}
\usepackage{amssymb}
\usepackage{amsfonts,amsbsy}
\usepackage{amsmath}
\usepackage{dcolumn}% Align table columns on decimal point
\usepackage{bm}
\usepackage{pifont}

\usepackage{graphicx}
\usepackage{amssymb}
\usepackage{epstopdf}

\usepackage[FIGTOPCAP,nooneline]{subfigure}

\usepackage[latin1]{inputenc}
\usepackage{amsfonts,amsthm,amssymb,amsxtra,amscd,latexsym}

\begin{document}

%\bibliographystyle{apsrev}

%%%%%%%%%%%%%%%%%%%%%%%%%%%%%%%%%%%%%%%%%
\title{Boundary-induced orientation of dynamic filament networks and vesicle agglomerations}
%\title{Particle accumulation driven by boundary-induced polarity of dynamic filament networks} 
%\title{Boundary-induced polarity of dynamic filament networks: Effects on transport properties}
%%%%%%%%%%%%%%%%%%%%%%%%%%%%%%%%%%%%%%%%%

\author{Philip Greulich}%
\affiliation{%
SUPA, School of Physics and Astronomy, University of Edinburgh, Edinburgh, UK
}%
\affiliation{%
Fachrichtung Theoretische Physik, Universit\"at des Saarlandes, Saarbr\"ucken, Germany}%
% \email{pg@thp.uni-koeln.de}

\author{Ludger Santen}%
% \email{as@thp.uni-koeln.de}
\affiliation{%
Fachrichtung Theoretische Physik, Universit\"at des Saarlandes, Saarbr\"ucken, Germany}%

\date{\today}% It is always \today, today,
             %  but any date may be explicitly specified

\begin{abstract}
We find a statistical mechanism that can adjust orientations of intracellular filaments to cell geometry in absence of organizing centers. The effect is based on random and isotropic filament (de-)polymerization dynamics and is independent of filament interactions and explicit regulation. It can be understood by an analogy to electrostatics and appears to be induced by the confining boundaries; for periodic boundary conditions no orientational bias emerges. Including active transport of particles, the model reproduces experimental observations of vesicle accumulations in transected axons.

%We investigate active transport of particles on filament networks that are evolving by stochastic dynamics, motivated by intracellular filaments. The intrinsic filament dynamics are isotropic and homogeneous such that the network structure is random and unbiased for periodic boundary conditions. We show that confining boundary conditions are leading to a  self-organized orientation of the network, even without explicit regulation and interactions. The orientation can lead to separation of particle species adjusting to the enclosing geometry. The underlying mechanism can be understood by a linear theory analogue to electrostatics. Applied to a cylindric geometry the model reproduces vesicle agglomerations and separation observed in axons after transection. This agglomeration mechanism, in fact, plays an important role in neuronal regeneration after mechanical damages.
\end{abstract}

\maketitle

%%%%%%%%%%%%%%%%%%%%%%%%%%%%%%%%%%%%%%%%%%%%%%%%%%%%%%%%%%%%%%%%%%%%%%%%%%%%%%%%%%%%%%%%

In living cells, cargo like nutrients, proteins or organelles, has to be carried to distinct locations in the cell. Many of these objects are transported by motor proteins (kinesin, dynein and myosin) that bind cargo and move along polarized intracellular filaments (microtubules, actin filaments) in a directed manner \cite{alberts}. The direction of movement is determined by the polarity of the filament (\emph{plus-} and a \emph{minus-}end) and the motor protein species. 
%The directed dynamics of this process is driven by the input of energy (provided by nucleotides) and is intrinsically a non-equilibrium process. 

For particular cell functions, accumulations of cargo and concentration gradients are needed at distinct locations of the cell. 
%In case of motor driven intracellular transport complex patterns emerge, which are built by the interplay between the dynamics of intracellular network and different proteins, which regulate the cellular structure.
For that purpose the arrangement of filaments, which determines the distribution of cargo, has to be organized properly. In some cases filaments are organized by auxiliary components to form globally polarized structures (e.g. the microtubule organizing center \cite{boehm_MTorg}). Next to these highly regulated mechanisms, motor driven transport processes take place also on less organized networks. 
In this article we consider a scenario where the initially ordered network of microtubules (\emph{MT}s) in a neuronal axon is destroyed after transection of the axon. Experimental observations show that after transection, vesicles agglomerate adjusting to the geometry of the cut axon \cite{spira_trap,spira_trap2}. This vesicle agglomeration is important for the formation of a \emph{growth cone} that is needed to recover neuronal connections. It was supposed that a re-organization of the microtubule network creates effective kinetic traps for the vesicles.

Mechanisms for self-organization of intracellular filaments were proposed in previous works, relying on extrinsic chemical gradients induced by external signals \cite{maly_actin_selforg, schaus_actin_selforg}, or spontaneous structure formation driven by the effective interactions of filaments mediated by molecular motors \cite{kruse_actinasters,kruse_juelicher_actin_selforg_1,kruse_juelicher_actin_selforg_2,astrom_filaments+motors,hawkins_voituriez_polarization}.  

Here, we use a complementary approach: We ask the question how structures self-organize without relying on explicit regulation mechanisms (e.g. by pre-established chemical gradients) or interactions, and discuss if they are necessary to explain the alignment of filament orientations and subsequent vesicle agglomerations in finite volumes. 
We consider explicitly a simple homogeneous and isotropic stochastic model for filament growth dynamics that neglects filament-filament interactions  \cite{celltransport_per}. The model considers basic filament nucleation and (de-) polymerization dynamics as considered in previous models (e.g.  \cite{microtubule_mathmodel}), without an a priori preferred direction or position of filaments nor any gradients. However, the network and particle dynamics are assumed to be confined to a given geometry corresponding to the finite cell volume.  We show that the confining boundary conditions in fact induce an orientation of the filament ensemble. 
We also present a linear theory for filament network evolution which is analogous to electrostatics and able to explain the observed orientation of filaments.

 %%%%%%%%%%%%%%%%%%%%%%%%%%%%%%%%%%%%%%%%%%%%%%%%%%%%%%%%%%%%%%%%%%%%%%%%%%%%%%%%%%%%

Our model is defined analogue to \cite{celltransport_per}. We apply stochastic dynamics in continuous space for filament and particle dynamics. The model rules capture the basics of filament nucleation and (de-)polymerization dynamics \cite{alberts}: Filaments are nucleated with spatially homogeneous rate $\omega_n\rho_{mon}\rho_{nuc}$ and isotropic random orientation. Here the values $\rho_{mon}$ and $\rho_{nuc}$ denote the (conserved) quantity of monomers and nucleation seeds in the cytosol. Filaments can grow by including segments of length $d_s$ at their \emph{plus-}end with rate $\omega_g\rho_{mon}$ and shrink by dissociating a segment at the \emph{minus-} end with rate $\omega_s$. After nucleation, the minus-end remains \emph{capped} for some time such that depolymerization at the minus-end is not possible. This restriction is removed with uncapping rate $\omega_{u}$  (which is non-zero only for actin filaments).
% such that polymerization at the plus-end and depolymerization at the minus end drive \emph{treadmilling} \cite{alberts}.

\begin{figure}
 \begin{center}
 \subfigure[]{\resizebox{0.4\columnwidth}{!}{\includegraphics{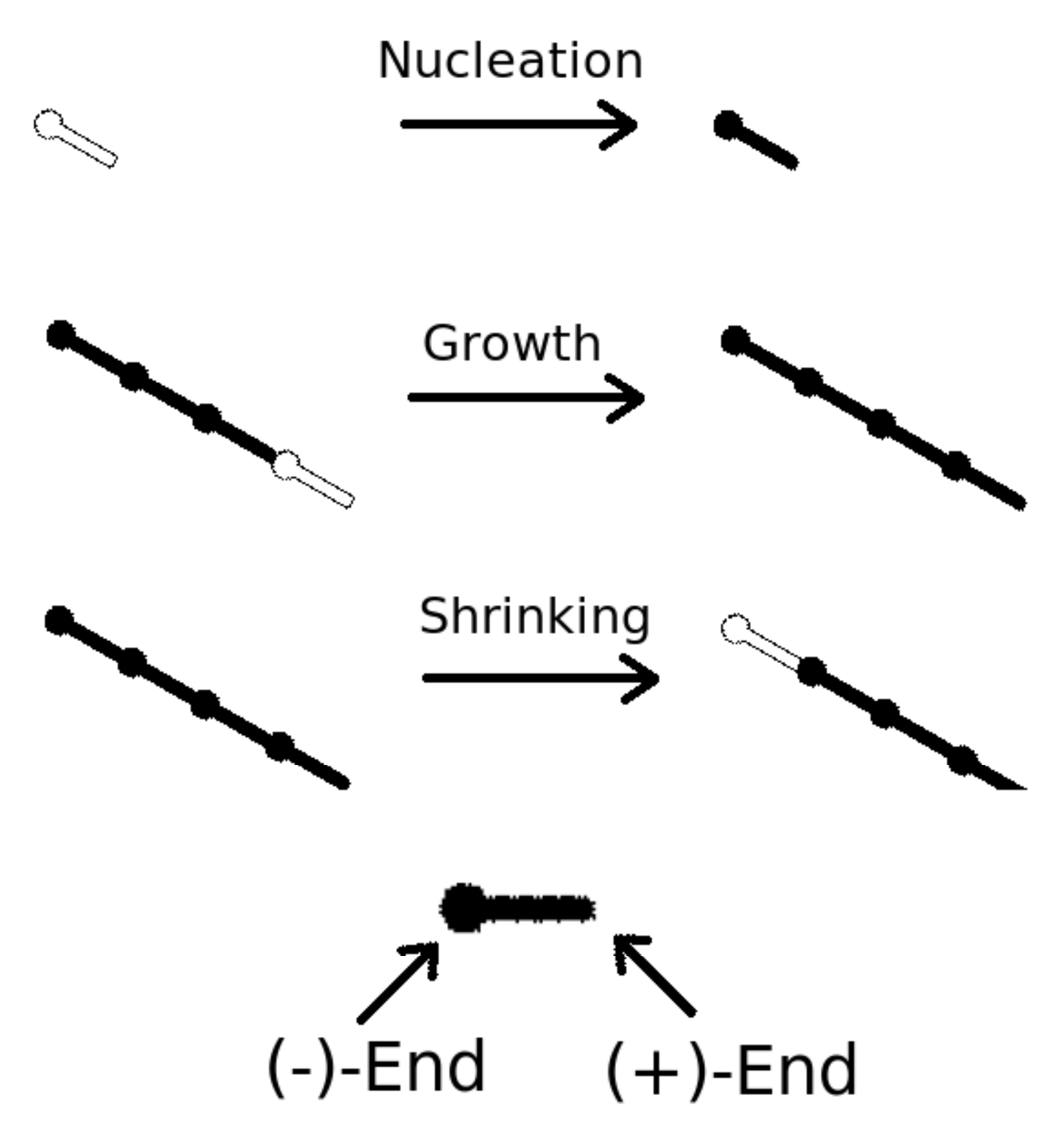}}}
  \subfigure[]{\resizebox{0.5\columnwidth}{!}{\includegraphics{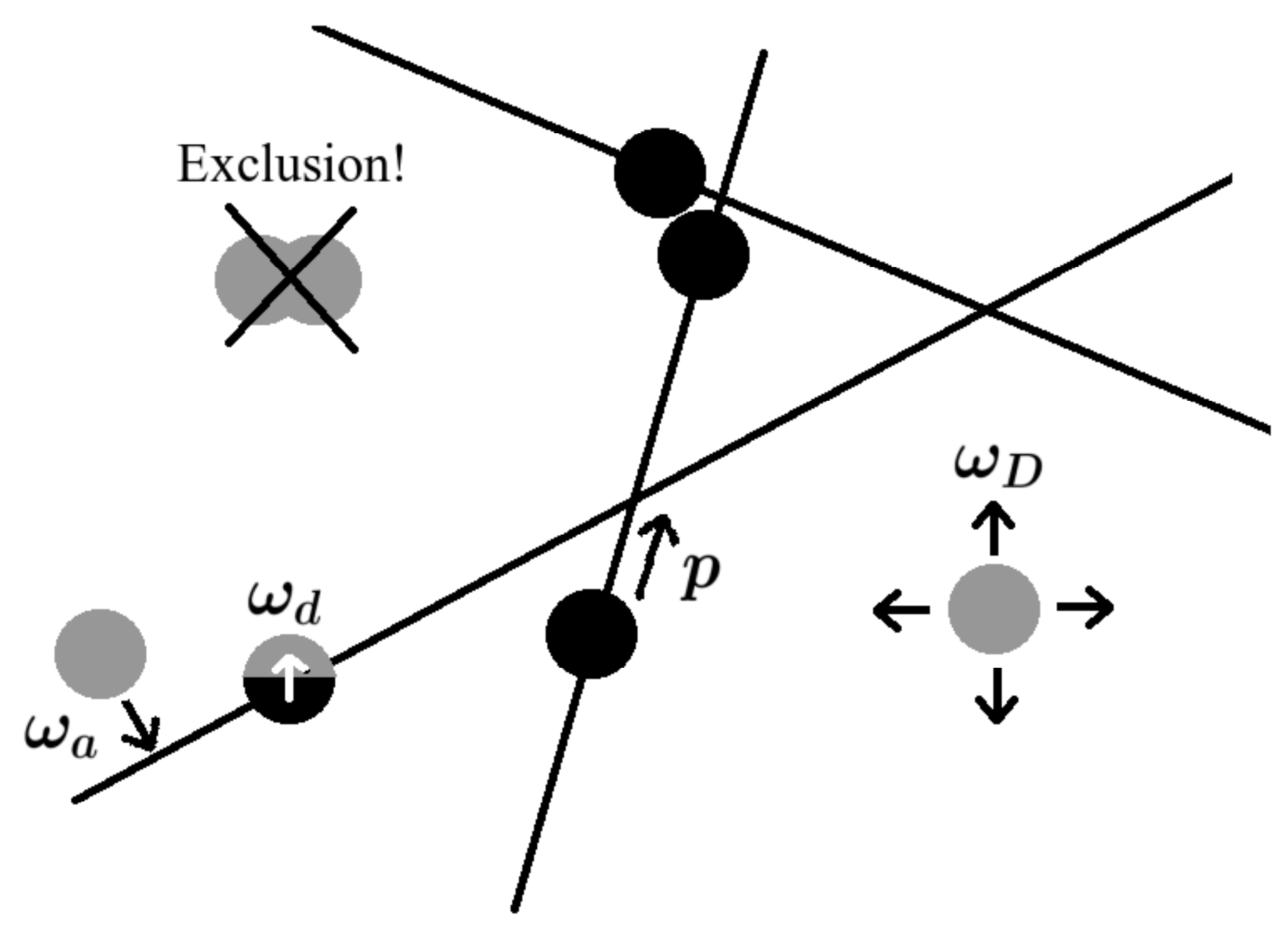}}}
 \caption{\label{illust_filament_dynamics} Illustration of model dynamics. (a) Filaments are implemented as sequences of linear segments. These are polarized, with a plus-end where segments are created to elongate, and a minus-end where segments dissociate causing shrinking. (b) Dark gray discs are detached particles, black discs represent particles attached to filaments. %stepping to adjacent subunits (distance $d_s$) with rate $p$. Particles can attach to filaments with rate $\omega_a$ if they are within the binding distance $d_b$ and detach with rate $\omega_d$. Overlapping is inhibited due to exclusion.
}
 \end{center}
 \end{figure}

Particles are implemented as hard-core spheres with radius $r_{p}$ exhibiting mutual steric exclusion.  \emph{Detached} particles perform a continuous random walk in space. If they are within the binding range of a filament $d_b$, they can \emph{attach} to the filament (rate $\omega_a$) and perform directed motion until they \emph{detach} with rate $\omega_d$. We consider two species of particles: \emph{plus-particles} move towards the plus-end of the filament with rate $p$ per segment, while \emph{minus-particles} move to the minus ends. 
The model dynamics are illustrated in Fig. \ref{illust_filament_dynamics}. For a more detailed definition we refer to the supplementary material and \cite{celltransport_per}.

 %%%%%%%%%%%%%%%%%%%%%%%%%%%%%%%%%%%%%%%%%%%%%%%%%%%%%%%%%%%%%%%%%%%%%%%%%%%%%%%%%
 %%%%%%%%%%%%%%%%%%%%%%%%%%%%%%%%%%%%%%%%%%%%%%%%%%%%%%%%%%%%%%%%%%%%%%%%%%%%%%%%%%%%

%%%%%%%%%%%%%%%%%%%%%%%%%%%%%%%%%%%%%%%%%%%%%%%%%%%%%%%%%%%%%%%%%%%%%%%%%%%%%%%%%
\begin{figure}
\begin{center}
% \subfigure[]{\resizebox{0.7\columnwidth}{!}{\includegraphics{yeast_normal.pdf}}}
% \subfigure[]{\resizebox{0.7\columnwidth}{!}{\includegraphics{yeast_mutant.pdf}}}
% \subfigure[]{\resizebox{0.47\columnwidth}{!}{\includegraphics{config_bud_wa=0,25_3D.pdf}}}
% \subfigure[]{\resizebox{0.47\columnwidth}{!}{\includegraphics{config_bud_wa=0_3D.pdf}}}
% \subfigure[]{\includegraphics[width=0.8\columnwidth]{axon_dyn_illust.pdf}}
 \subfigure[]{\includegraphics[width=0.75\columnwidth]{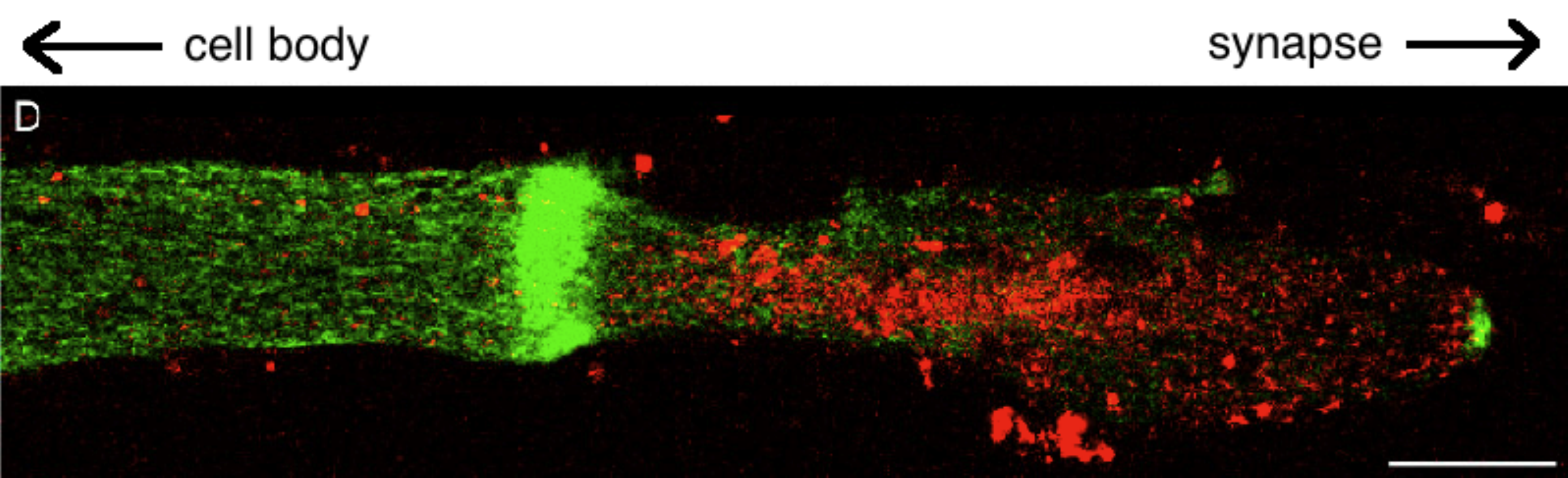}}
 %\subfigure[]{\includegraphics[width=0.7\columnwidth]{yeast_mutant.pdf}}

% \subfigure[]{\includegraphics[width=0.8\columnwidth]{axon_illust.pdf}}
 \subfigure[]{\includegraphics[width=0.75\columnwidth]{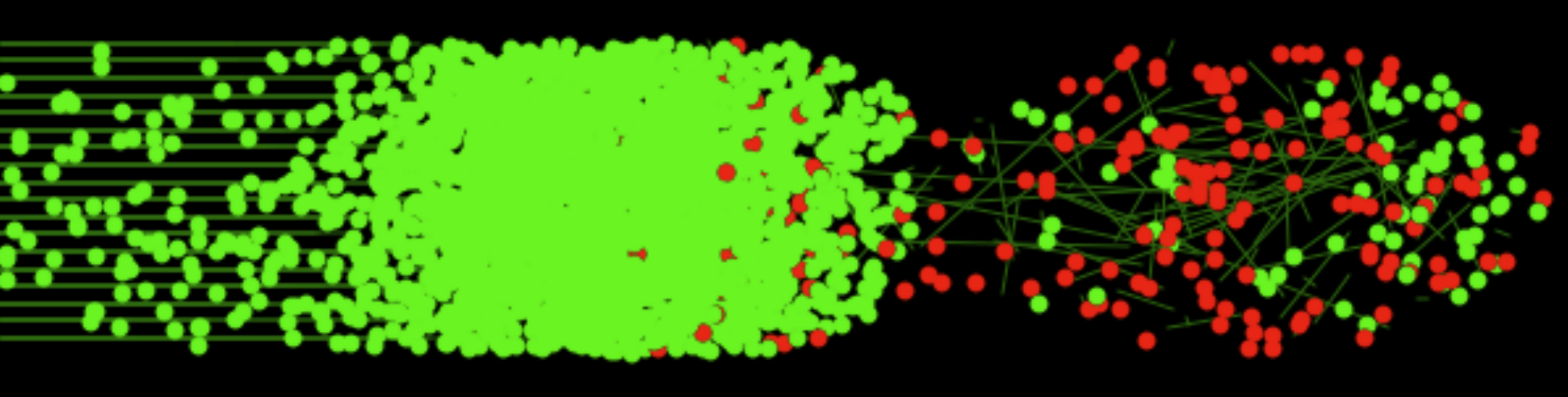}}
 %\subfigure[]{\includegraphics[width=0.45\columnwidth]{config_bud_wa=0_3D.pdf}}
\caption{\label{axon_fig} Comparison between experimental results of Erez et al. \cite{spira_trap} (a) and model results (b). The same color-coding is used: anterograde (plus-) particles are green, retrograde (minus-) particles are red. The shape of the model axon is based on the experiment (bar=15 $\mu m$). The deviations from the cylindrical shape have only little influence on the particle distribution, hence the accumulation is \emph{not} result of a (diffusive) bottleneck (see supplementary material).}
\end{center}
\end{figure}
%%%%%%%%%%%%%%%%%%%%%%%%%%%%%%%%%%%%%%%%%%%%%%%%%%%%%%%%%%%%%%%%%%%%%%%%%%%%%%%%%%%%

We want to apply the model to the experimental setup studied in \cite{spira_trap}:  Healthy axons usually contain bundles of axially orientated microtubules with plus ends pointing towards the synapse. In the experiments, axons were transected leading to dissolution of microtubules near the location of the cut, due to calcium influx (right part of Fig. \ref{axon_fig}(a)). In that region, microtubules subsequently reassemble randomly in absence of an organizing center. Non-destroyed MTs remain and can grow again. One observes an accumulation of plus- (anterograde) vesicles at the tip and at some distance from the tip as well as minus- (retrograde) vesicles between these locations. 

In the model we implement this scenario by introducing a confining volume that is based on the cylindrical shape of an axon. We use an idealized cylindrical shape as well as a shape similar to the contour of the axon used in the experiment of Erez et al. \cite{spira_trap}. The finite volume leads to a straightforward modification of the filament dynamics: Filaments are not allowed to grow or nucleate outside the volume (see Fig. \ref{axon_fig}(b)).  Starting from parallel bundles of microtubules, they get dissolved (transection) within some distance from the tip (right half of the figure) and new filaments reassemble randomly without preferred direction, according to the rules defined above. Moreover, we assume that filament growth is inhibited by the steric hard-core particles such that filaments are not allowed to grow through a particle. In addition, we implement microtubule catastrophe-rescue dynamics following the model in  \cite{microtubule_mathmodel}.
%: filaments that hit a boundary can turn into a plus-end shrinking state state (\emph{catastrophe}) with rate $\omega_c$ and can resume the initial growing state with a given rate $\omega_r$. 
At the left boundary, plus particles enter the system with rate $\alpha_1$ where also minus particles exit the system with rate $\beta_2$. At the other boundaries (representing the membrane), plus particles exit (\emph{exocytosis}) with rate $\beta_1$ and minus particles enter the system (\emph{endocytosis}) with rate $\alpha_2$. 
%(See the supplementary material for parameter values.).  

In Fig. \ref{axon_fig}(b) a particle configuration 24000 time steps after transection (corresponding to about 4 min in real time) is displayed. We used the biologically motivated parameters as discussed in the supplementary material. Starting with parallel filaments, they are dissolved in the right part of the system after 7000 time steps, whereupon random nucleation and growth of filaments, according to the dynamics discussed above, takes place in this region.

%%%%%%%%%%%%%%%%%%%%%%%%%%%%%%%%%%%%%%%%%%%%%%%%%%%%%%%%%%%%%%%%%%%%%%%%%%%%%%%%%
\begin{figure}
\begin{center}
\resizebox{0.53\columnwidth}{!}{\includegraphics[angle=270]{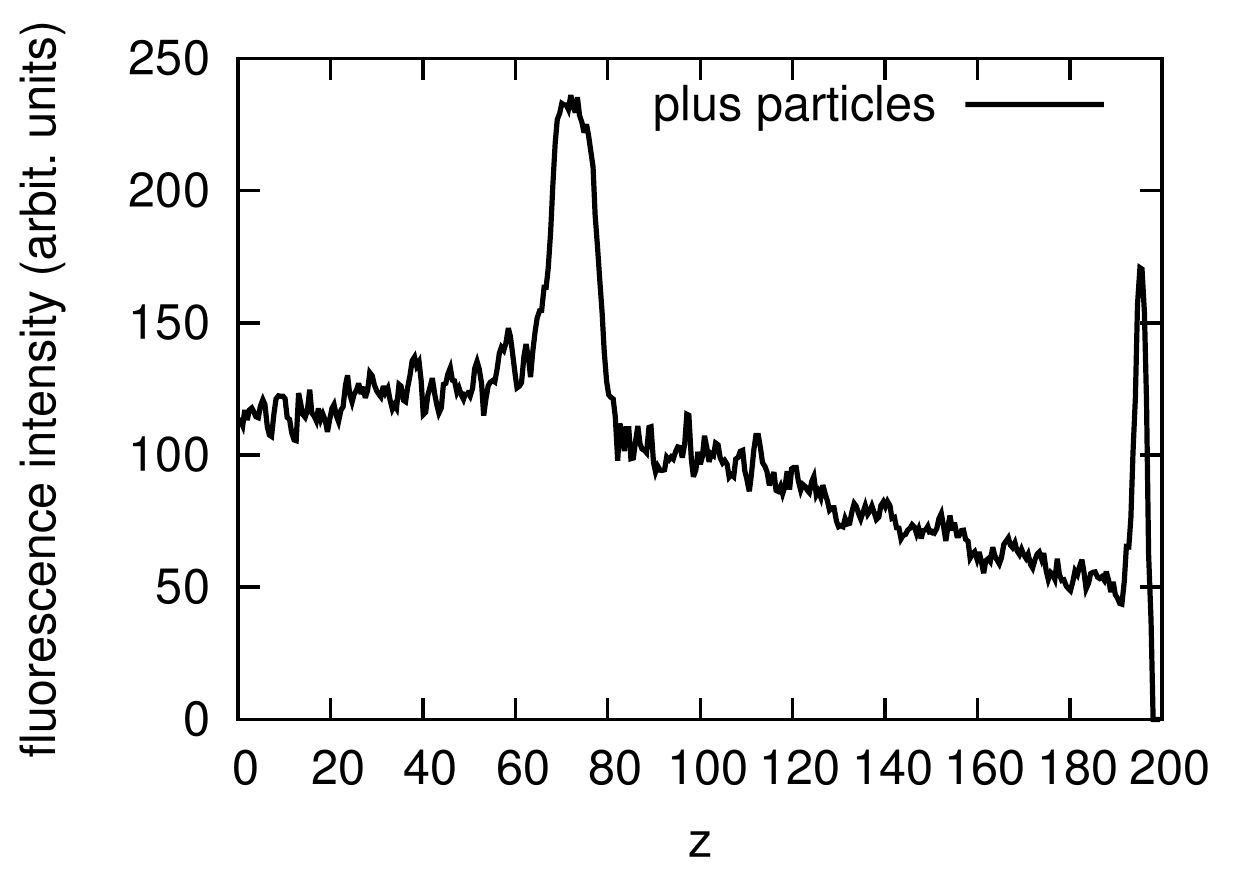}}
\resizebox{0.57\columnwidth}{!}{\includegraphics[angle=270]{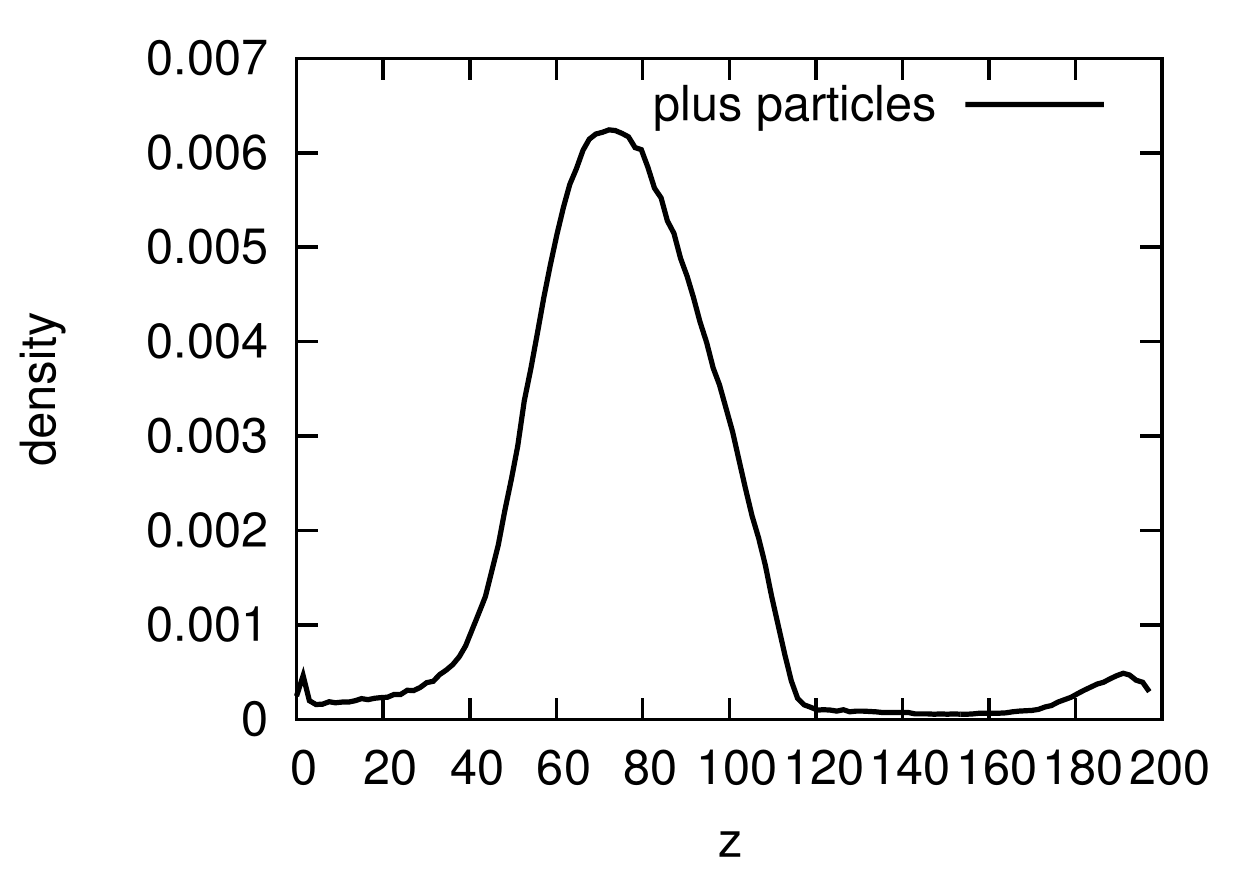}}
\caption{\label{trap_fig} Comparison between experimental \cite{spira_trap}(top) and simulated particle distribution (bottom). z = cylinder axis coordinate.}
\end{center}
\end{figure}
%%%%%%%%%%%%%%%%%%%%%%%%%%%%%%%%%%%%%%%%%%%%%%%%%%%%%%%%%%%%%%%%%%%%%%%%%%%%%%%%%%%%

The comparison between experimental and model results shows that the model is able to reproduce the structure of the plus particle traps as well in the middle of the system and at the tip, but it underestimates its strength (Fig. \ref{trap_fig}(b)). 
%This will be discussed later.

In order to understand the emergence of particle accumulations, we develop a linear theory for the network structure. The orientations of filament segments will be denoted by a unit vector $\mathbf d$ describing the direction of the plus-end. We quantify the structure of the network by a vector field \begin{equation}
\mathbf F(\mathbf x):= \left\langle\lim_{v\to 0} \frac{1}{v} \sum_{\mbox{fil. in v}} \mathbf d(\mathbf x) \right\rangle =  \rho_s \langle \mathbf d  (\mathbf x) \rangle
\end{equation}
which we call \emph{filament field}. Here, $v$ is a volume containing the point $\mathbf x$. The sum includes all filament segments in $v$, $\rho_s$ denotes the density of filament segments at the point $\mathbf x$ and $\left\langle\cdots \right\rangle$ denotes the ensemble average. The direction of $\mathbf F$ represents the average orientation of filaments, whereas its amplitude is determined by the density of filaments. $\mathbf F(\mathbf x)$ results from a linear superposition of filament orientations at the point $\mathbf x$ which allows to express it by a Green's function as
\begin{equation}
\label{greens_formalism}
\mathbf F(\mathbf x)=\int_{V} {\mathbf G}(\mathbf x-\mathbf x') \rho_n(\mathbf x') \,d{x'}^3
\end{equation}
where ${\mathbf G}(\mathbf x-\mathbf x')$ is the field if nucleation of filaments would only be possible in a single point $\mathbf x'$, and $\rho_n(\mathbf x)$ denotes the relative nucleation rate. We consider the dynamics to be spatially homogeneous within the volume $V$ where filaments can nucleate, hence  $\rho_n(\mathbf x)=1$ within $V$ and $\rho_n(\mathbf x)=0$ outside \footnote{This normalization is chosen to separate the influence of dynamics (in $\mathbf G(\mathbf x-\mathbf x')$ ) and geometry (in $\rho_n(\mathbf x)$) }. However, (\ref{greens_formalism}) can also be applied, when nucleation dynamics vary spatially.
 
Since the orientation of filaments always points away from the nucleation point, the Green's function $\mathbf G(\mathbf x)$, which considers only nucleation in the origin, must have a radial structure $\mathbf G(\mathbf x)= G(\mathbf x) \mathbf e_r$. In fact, $G(\mathbf x)$ is the probability that a filament is within a volume unit $dx^3$ around $\mathbf x$. Since filaments are not correlated, this can be expressed by the probability $P(\mathbf x)$ that the filament is directed  towards $\mathbf x$ and, assuming ergodicity, the average time $\tau(\mathbf x)$ the filament is present in $dx^3$: $G(\mathbf x)= P(\mathbf x)\tau(r)$. Because a filament grows isotropically in arbitrary direction, the probability that it passes the volume $dx^3$ located at distance $r=|\mathbf x|$ from the origin is given by $P(\mathbf x) = P(r) \propto 1/r^{2}$. In the stationary state, the net polymerization rate $\omega_g\rho_{\rm mon}$ must equal the depolymerization rate $\omega_s$ due to conservation of monomers. If we neglect shrinking at the plus end, 
%(e.g. for actin filaments) 
a filament can only disappear at a given point by minus-end depolymerization. The average filament length $l$ is independent of the distance $r$ and hence the time the filament is present at $\mathbf x$ is $\tau = l / \omega_s$ which is also independent of $r$. 

After all, the overall value of the filament field is 
\begin{equation}
\label{fil_field_coulomb}
{\mathbf G(r)} \propto \frac{\mathbf e_r}{r^{2}} \,\,\, .
\end{equation}
This form corresponds to the Green's function of electrostatics (Coulomb law). As a result, the filament field inside a volume $V$ with homogeneous nucleation rate corresponds to a electrostatic field of a homogeneous charge distribution within $V$. 
 
In order to test these results quantitatively, we consider more simple boundary conditions than in the previous example. Specifically we apply spherical boundary conditions (sbc), where the dynamic filaments are confined to a sphere of radius $R$. For comparison we also consider periodic boundary conditions (pbc). 

%%%%%%%%%%%%%%%%%%%%%%%%%%%%%%%%%%%%%%%%%%%%%%%%%%%%%%%%%%%%%%%%%%%%%%%%%%%%%%%%%
\begin{figure}
\begin{center}
\resizebox{0.55\columnwidth}{!}{\includegraphics{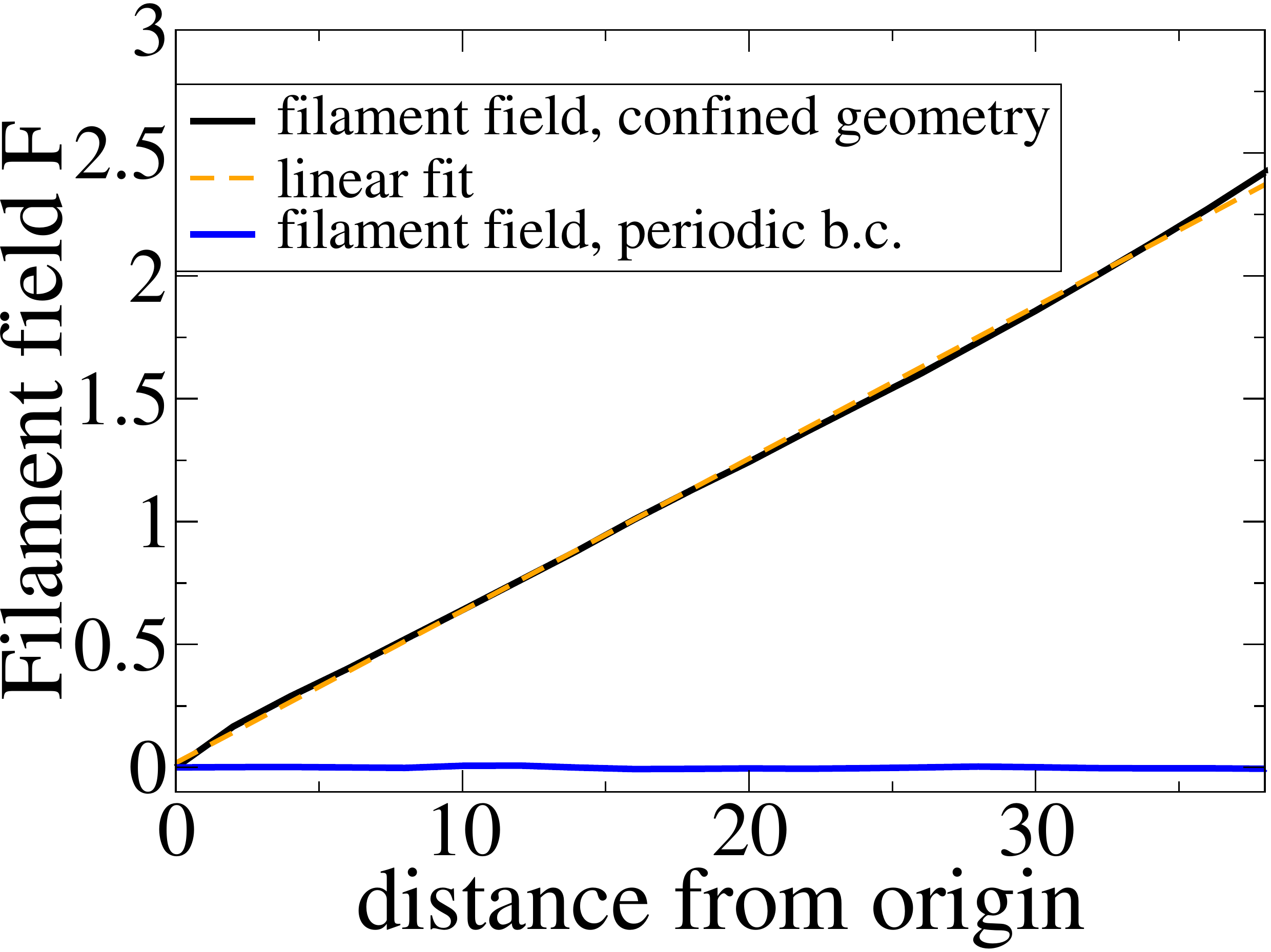}}
\caption{\label{filfield} Radial component of the filament field $F(r)$ for default parameters in 3D. For pbc there is no bias in the network structure as expected due to translational symmetry. For confining sbc, one observes approximately a linear behavior as predicted by the analytical theory.}
\end{center}
\end{figure}
%%%%%%%%%%%%%%%%%%%%%%%%%%%%%%%%%%%%%%%%%%%%%%%%%%%%%%%%%%%%%%%%%%%%%%%%%%%%%%%%%%%%

Due to translational symmetry for pbc and non-biased dynamics there is supposed to be no overall bias and the field $\mathbf F$ vanishes. This is reproduced by simulations as can be seen in Fig. \ref{filfield}. However, if the filaments are confined to a sphere, the theory predicts that $\mathbf F$ has the same form as an electrostatic field inside a homogeneously charged sphere. For that geometry, Gauss' divergence theorem yields
\begin{equation}
\mathbf F(\mathbf x) \sim r \mathbf e_r \hspace{5mm} \mbox{for } r<R \,\,\, .
\end{equation}
This linear behavior of $\mathbf F$, with average radial orientation of filament plus ends towards boundaries, is indeed reproduced by simulations of filament ensembles (Fig. \ref{filfield}). 

\begin{figure}
\begin{center}
%\subfigure[]{\resizebox{0.48\columnwidth}{!}{\includegraphics{config_nofilbound.pdf}}}
\subfigure[]{\resizebox{0.41\columnwidth}{!}{\includegraphics{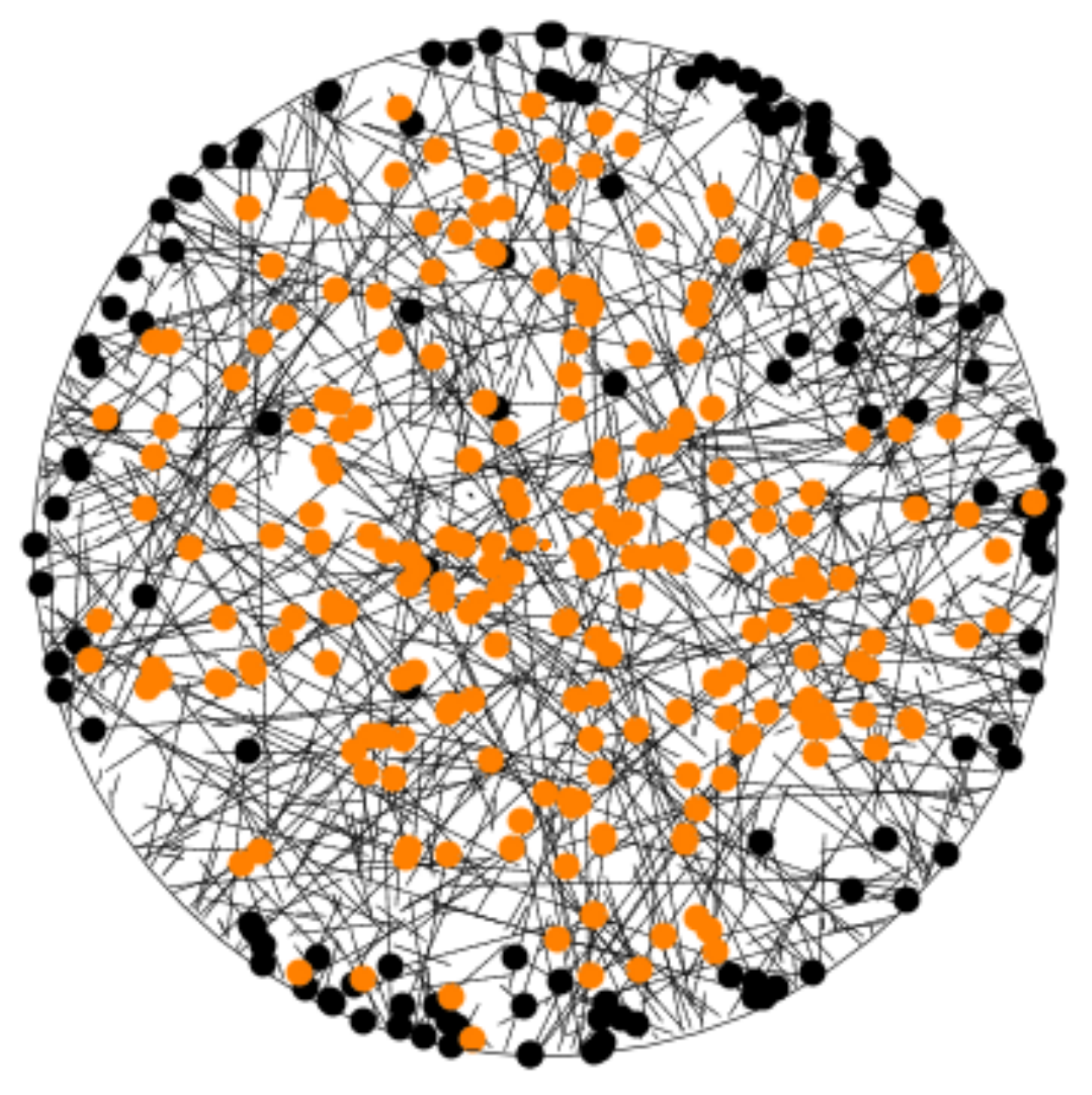}}}
\subfigure[]{\resizebox{0.54\columnwidth}{!}{\includegraphics{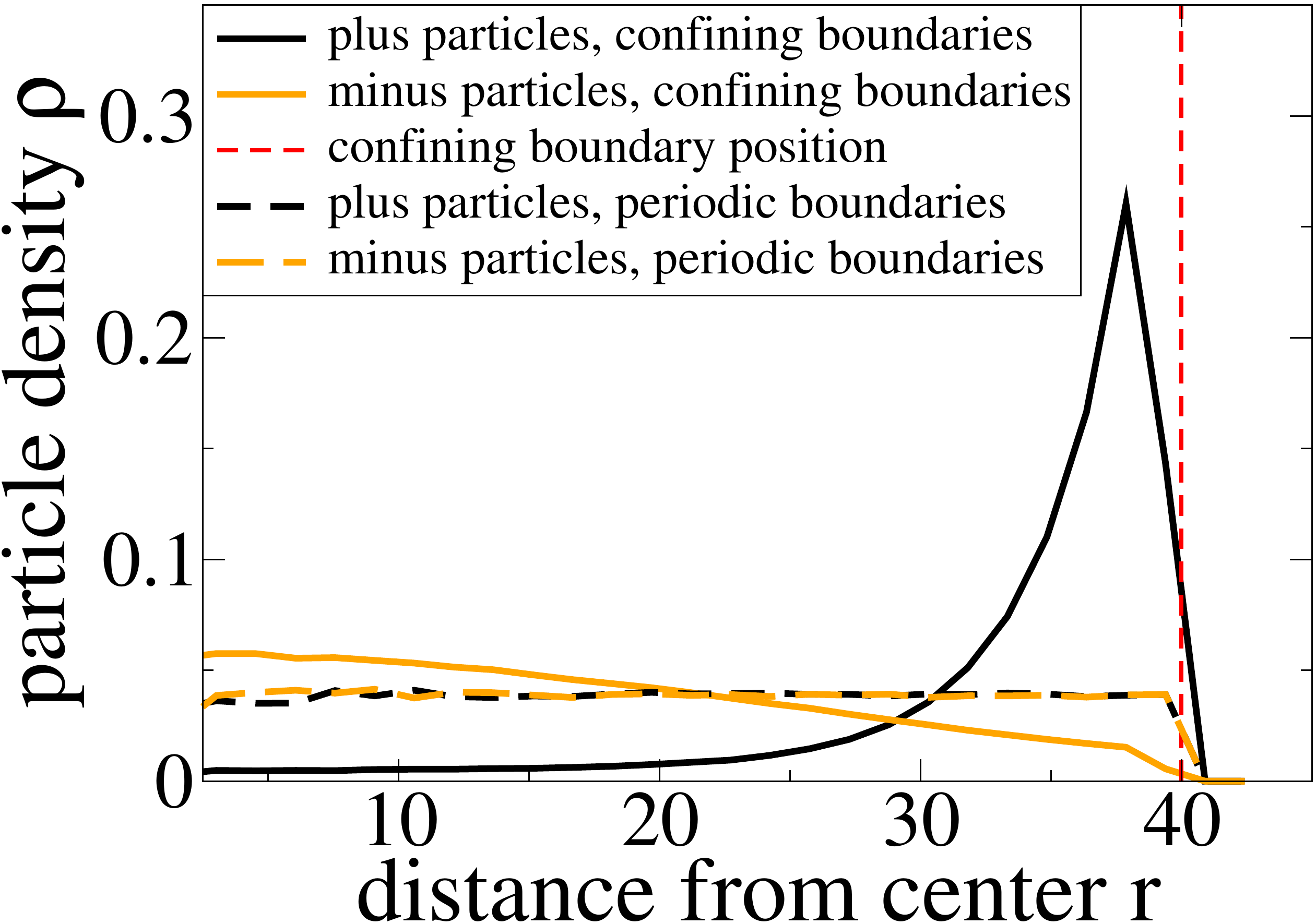}}}
\caption{\label{config_bound} Distributions of plus- (black) and minus-particles (orange) for spherical boundary conditions.(a) Snapshot of particle configurations (Slice plane).  (b) Radial density distribution of vesicles.}
\end{center}
\end{figure}

Particles follow the emerging radial bias of the filaments leading to a separation of particle species (Fig. \ref{config_bound}). Therefore, in the case of spherical boundary conditions, plus-particles accumulate at the boundaries while minus-particles are depleted at the boundaries and have a slight tendency towards the center. Analogue to electrostatics we conclude that, for an arbitrary geometry, random nucleation dynamics leads, in general, to an average bias of plus ends towards the boundaries. 

%%%%%%%%%%%%%%%%%%%%%%%%%%%%%%%%%%%%%%%%%%%%%%%%%%%%%%%%%%%%%%%%%%%%%%%%%%%%%%%%%
%%%%%%%%%%%%%%%%%%%%%%%%%%%%%%%%%%%%%%%%%%%%%%%%%%%%%%%%%%%%%%%%%%%%%%%%%%%%%%%%%%%%

However, for MT dynamics the Green's function needs some modifications. Since there is a chance for plus-end depolymerization, while minus ends are fixed, the Green's function has a finite range scale, given by the mean filament length. For larger distances, the effect of a nucleating filament is 'screened'. This means that the bias effect is only strong within a distance from the boundary that corresponds to the average filament length. Nonetheless, the \emph{direction} of the bias remains unaffected by the screening.  For our model geometry, the mean filament length ($80\, r_p$) is larger than the cylinder radius ($40\, r_p$), such that screening effects can be neglected. 

%%%%%%%%%%%%%%%%%%%%%%%%%%%%%%%%%%%%%%%%%%%%%%%%%%%%%%%%%%%%%%%%%%%%%%%%%%%%%%%%%
\begin{figure}
\begin{center}
\resizebox{0.9\columnwidth}{!}{\includegraphics{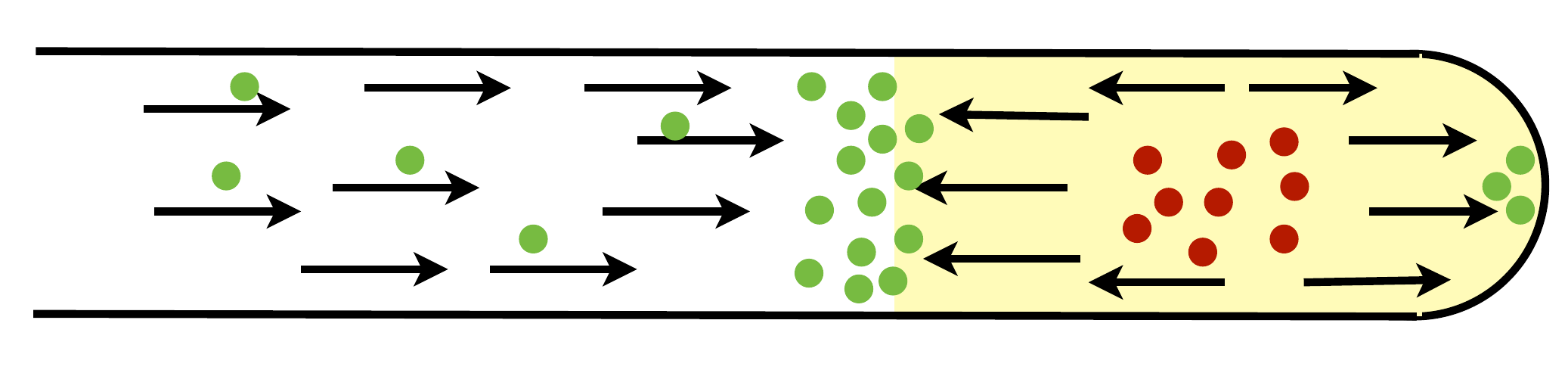}}
\caption{\label{axon-scheme_edit} Illustration of the re-organization of the microtubule network after transection. After dissolution of filaments, they re-assemble randomly in the right part (yellow region). According to our theory, plus ends of MTs (black arrows) are pointed towards the boundaries of this region. Adapted from \cite{spira_trap}}.
\end{center}
\end{figure}
%%%%%%%%%%%%%%%%%%%%%%%%%%%%%%%%%%%%%%%%%%%%%%%%%%%%%%%%%%%%%%%%%%%%%%%%%%%%%%%%%%%%

Within this theory, we can also explain the observed distribution of particles in case of a transected axon: After dissolution of MTs, random nucleation and growth induces a bias of plus ends outwards of the nucleating region, arranging (schematically) as in Fig. \ref{axon-scheme_edit} (only right part (yellow) is nucleating region). Together with the surviving microtubules at the left end, pointing to the right, this yields a trap for plus particles. In addition, a trap at the right tip emerges due to MTs pointing to the right. Our model, however, underestimates the strength of the particle traps compared to the experiments (Figs. \ref{axon_fig} and \ref{trap_fig}). This discrepancy supports the interpretation given in \cite{spira_trap} that interactions between filaments  (which we have neglected)  indeed increase orientational correlations to enhance structural inhomogeneities in the filament networks \cite{kruse2005} that stabilize and amplify the particle trapping. The bias predicted by our model thus can trigger an interaction-driven self-organization of the filaments towards the observed patterns.

To summarize, we studied a model for active transport on filament networks that evolve by isotropic and spatially homogeneous stochastic filament dynamics.
We showed that if filament dynamics is confined within a finite volume, the filaments become oriented following the geometry of the system, despite the absence of filament-filament interactions and chemical gradients. Applied to the geometry of a transected axon, this random dynamics is sufficient to explain the structure of experimentally observed vesicle traps, which is essential for axonal regeneration.
 
 We developed a linear theory which is analogous to electrostatics and describes correctly the alignment of filament orientations. This theory is assumed to be generic for arbitrary geometries and predicts a preferred orientation of filament plus-ends towards boundaries. 
Hence we have shown that in confined geometries, concentration gradients can generally be induced by randomly generated networks of polarized filaments, even without interaction-driven self-organization or external gradients.

Our results indicate that the geometry of the cell may have strong influence on the development of the structure of the filament network: Although our model results do not reproduce the strength of vesicle traps, their structure is well described. Therefore we conclude that the mechanism presented in this work may select the emerging structures of  interacting filament systems \cite{kruse_juelicher_actin_selforg_2,kruse_actinasters,astrom_filaments+motors,lee_filament_selforg}. This assumption is also supported by experiments providing evidence that the cell shape may also control the organization of the cytoskeleton \cite{terenna}. 

We want to thank M.E. Spira, Davide Marenduzzo and Marc Neef for fruitful discussions and Maren Westkott for image analysis. We also thank DFG Grants GK1276/1 and GZ SA864/3-1 for financial support. P.G. is postdoctoral fellow of DAAD.

%\bibliography{library}

\begin{thebibliography}{12}
\expandafter\ifx\csname natexlab\endcsname\relax\def\natexlab#1{#1}\fi
\expandafter\ifx\csname bibnamefont\endcsname\relax
  \def\bibnamefont#1{#1}\fi
\expandafter\ifx\csname bibfnamefont\endcsname\relax
  \def\bibfnamefont#1{#1}\fi
\expandafter\ifx\csname citenamefont\endcsname\relax
  \def\citenamefont#1{#1}\fi
\expandafter\ifx\csname url\endcsname\relax
  \def\url#1{\texttt{#1}}\fi
\expandafter\ifx\csname urlprefix\endcsname\relax\def\urlprefix{URL }\fi
\providecommand{\bibinfo}[2]{#2}
\providecommand{\eprint}[2][]{\url{#2}}

\bibitem[{\citenamefont{Alberts et~al.}(2002)}]{alberts}
\bibinfo{author}{\bibfnamefont{B.}~\bibnamefont{Alberts}} \bibnamefont{et~al.},
  \emph{\bibinfo{title}{{Molecular Biology of the Cell}}}
  (\bibinfo{publisher}{Garland}, \bibinfo{year}{2002}).

\bibitem[{\citenamefont{Greulich and Santen}(2010)}]{celltransport_per}
\bibinfo{author}{\bibfnamefont{P.}~\bibnamefont{Greulich}} \bibnamefont{and}
  \bibinfo{author}{\bibfnamefont{L.}~\bibnamefont{Santen}},
  \bibinfo{journal}{Eur. Phys. J. E} \textbf{\bibinfo{volume}{32}},
  \bibinfo{pages}{191} (\bibinfo{year}{2010}).

\bibitem[{\citenamefont{Alvarez and Zarour}(1983)}]{microtubule_areadens}
\bibinfo{author}{\bibfnamefont{J.}~\bibnamefont{Alvarez}} \bibnamefont{and}
  \bibinfo{author}{\bibfnamefont{J.}~\bibnamefont{Zarour}},
  \bibinfo{journal}{Exp Neurol.} \textbf{\bibinfo{volume}{79}},
  \bibinfo{pages}{283} (\bibinfo{year}{1983}).

\bibitem[{\citenamefont{Hinow et~al.}(2009)\citenamefont{Hinow, Rezania, and
  Tuszy\'nski}}]{microtubule_mathmodel}
\bibinfo{author}{\bibfnamefont{P.}~\bibnamefont{Hinow}},
  \bibinfo{author}{\bibfnamefont{V.}~\bibnamefont{Rezania}}, \bibnamefont{and}
  \bibinfo{author}{\bibfnamefont{J.~A.} \bibnamefont{Tuszy\'nski}},
  \bibinfo{journal}{Phys. Rev. E} \textbf{\bibinfo{volume}{80}},
  \bibinfo{pages}{31904} (\bibinfo{year}{2009}).

\bibitem[{\citenamefont{Yu and Baas}(1994)}]{microtubule_length}
\bibinfo{author}{\bibfnamefont{W.}~\bibnamefont{Yu}} \bibnamefont{and}
  \bibinfo{author}{\bibfnamefont{P.~W.} \bibnamefont{Baas}},
  \bibinfo{journal}{The Journal of Neuroscience} \textbf{\bibinfo{volume}{14}},
  \bibinfo{pages}{2818} (\bibinfo{year}{1994}).

\bibitem[{\citenamefont{Chen and Hill}(1988)}]{attachmentrate_hill_1}
\bibinfo{author}{\bibfnamefont{Y.-d.} \bibnamefont{Chen}} \bibnamefont{and}
  \bibinfo{author}{\bibfnamefont{T.~L.} \bibnamefont{Hill}},
  \bibinfo{journal}{Proc. Natl. Acad. Sci.} \textbf{\bibinfo{volume}{85}},
  \bibinfo{pages}{431} (\bibinfo{year}{1988}).

\bibitem[{\citenamefont{Erez et~al.}(2007)}]{spira_trap}
\bibinfo{author}{\bibfnamefont{H.}~\bibnamefont{Erez}} \bibnamefont{et~al.},
  \bibinfo{journal}{The Journal of Cell Biology}
  \textbf{\bibinfo{volume}{176}}, \bibinfo{pages}{497} (\bibinfo{year}{2007}).

\bibitem[{\citenamefont{Cassimeris et~al.}(1988)\citenamefont{Cassimeris,
  Pryer, and Salmon}}]{microtubule_param_3}
\bibinfo{author}{\bibfnamefont{L.}~\bibnamefont{Cassimeris}},
  \bibinfo{author}{\bibfnamefont{N.~K.} \bibnamefont{Pryer}}, \bibnamefont{and}
  \bibinfo{author}{\bibfnamefont{E.~D.} \bibnamefont{Salmon}},
  \bibinfo{journal}{The Journal of Cell Biology}
  \textbf{\bibinfo{volume}{107}}, \bibinfo{pages}{2223} (\bibinfo{year}{1988}).

\bibitem[{\citenamefont{Korgel et~al.}(1998)\citenamefont{Korgel, van Zanten,
  and Monbouquette}}]{vesicle_radius}
\bibinfo{author}{\bibfnamefont{B.~A.} \bibnamefont{Korgel}},
  \bibinfo{author}{\bibfnamefont{J.~H.} \bibnamefont{van Zanten}},
  \bibnamefont{and} \bibinfo{author}{\bibfnamefont{H.~G.}
  \bibnamefont{Monbouquette}}, \bibinfo{journal}{Biophysical Journal}
  \textbf{\bibinfo{volume}{74}}, \bibinfo{pages}{3264} (\bibinfo{year}{1998}).

\bibitem[{\citenamefont{Kwinter et~al.}(2009)\citenamefont{Kwinter, Lo, Mafi,
  and Silverman}}]{axon_vesicles}
\bibinfo{author}{\bibfnamefont{D.~M.} \bibnamefont{Kwinter}},
  \bibinfo{author}{\bibfnamefont{K.}~\bibnamefont{Lo}},
  \bibinfo{author}{\bibfnamefont{P.}~\bibnamefont{Mafi}}, \bibnamefont{and}
  \bibinfo{author}{\bibfnamefont{M.~A.} \bibnamefont{Silverman}},
  \bibinfo{journal}{Neuroscience} \textbf{\bibinfo{volume}{162}},
  \bibinfo{pages}{1001} (\bibinfo{year}{2009}).

\bibitem[{\citenamefont{Luby-Phelps}(2000)}]{vesicle_diff}
\bibinfo{author}{\bibfnamefont{K.}~\bibnamefont{Luby-Phelps}},
  \bibinfo{journal}{International Review of Cytology}
  \textbf{\bibinfo{volume}{192}}, \bibinfo{pages}{189} (\bibinfo{year}{2000}).

\bibitem[{\citenamefont{Mackey et~al.}(1981)\citenamefont{Mackey, Schuessler,
  Goldberg, and Schwartz}}]{microtubule_vesdens}
\bibinfo{author}{\bibfnamefont{S.}~\bibnamefont{Mackey}},
  \bibinfo{author}{\bibfnamefont{G.}~\bibnamefont{Schuessler}},
  \bibinfo{author}{\bibfnamefont{D.~J.} \bibnamefont{Goldberg}},
  \bibnamefont{and} \bibinfo{author}{\bibfnamefont{J.~H.}
  \bibnamefont{Schwartz}}, \bibinfo{journal}{Biophys J.}
  \textbf{\bibinfo{volume}{36}}, \bibinfo{pages}{455} (\bibinfo{year}{1981}).

\end{thebibliography}


\begin{thebibliography}{17}
\expandafter\ifx\csname natexlab\endcsname\relax\def\natexlab#1{#1}\fi
\expandafter\ifx\csname bibnamefont\endcsname\relax
  \def\bibnamefont#1{#1}\fi
\expandafter\ifx\csname bibfnamefont\endcsname\relax
  \def\bibfnamefont#1{#1}\fi
\expandafter\ifx\csname citenamefont\endcsname\relax
  \def\citenamefont#1{#1}\fi
\expandafter\ifx\csname url\endcsname\relax
  \def\url#1{\texttt{#1}}\fi
\expandafter\ifx\csname urlprefix\endcsname\relax\def\urlprefix{URL }\fi
\providecommand{\bibinfo}[2]{#2}
\providecommand{\eprint}[2][]{\url{#2}}

\bibitem[{\citenamefont{Alberts et~al.}(2002)}]{alberts}
\bibinfo{author}{\bibfnamefont{B.}~\bibnamefont{Alberts}} \bibnamefont{et~al.},
  \emph{\bibinfo{title}{{Molecular Biology of the Cell}}}
  (\bibinfo{publisher}{Garland}, \bibinfo{year}{2002}).

\bibitem[{\citenamefont{Wolf and B\"{o}hm}(1997)}]{boehm_MTorg}
\bibinfo{author}{\bibfnamefont{K.~W.} \bibnamefont{Wolf}} \bibnamefont{and}
  \bibinfo{author}{\bibfnamefont{K.~J.} \bibnamefont{B\"{o}hm}},
  \bibinfo{journal}{Biologie in unserer Zeit} \textbf{\bibinfo{volume}{27.2}},
  \bibinfo{pages}{87} (\bibinfo{year}{1997}).

\bibitem[{\citenamefont{Erez et~al.}(2007)}]{spira_trap}
\bibinfo{author}{\bibfnamefont{H.}~\bibnamefont{Erez}} \bibnamefont{et~al.},
  \bibinfo{journal}{The Journal of Cell Biology}
  \textbf{\bibinfo{volume}{176}}, \bibinfo{pages}{497} (\bibinfo{year}{2007}).

\bibitem[{\citenamefont{Kamber et~al.}(2009)\citenamefont{Kamber, Erez, and
  Spira}}]{spira_trap2}
\bibinfo{author}{\bibfnamefont{D.}~\bibnamefont{Kamber}},
  \bibinfo{author}{\bibfnamefont{H.}~\bibnamefont{Erez}}, \bibnamefont{and}
  \bibinfo{author}{\bibfnamefont{M.~E.} \bibnamefont{Spira}},
  \bibinfo{journal}{Experimental neurology} \textbf{\bibinfo{volume}{219}},
  \bibinfo{pages}{112} (\bibinfo{year}{2009}).

\bibitem[{\citenamefont{Maly and Borisy}(2001)}]{maly_actin_selforg}
\bibinfo{author}{\bibfnamefont{I.~V.} \bibnamefont{Maly}} \bibnamefont{and}
  \bibinfo{author}{\bibfnamefont{G.~G.} \bibnamefont{Borisy}},
  \bibinfo{journal}{Proc. Natl. Acad. Sci.} \textbf{\bibinfo{volume}{98}},
  \bibinfo{pages}{11324} (\bibinfo{year}{2001}).

\bibitem[{\citenamefont{Schaus et~al.}(2007)\citenamefont{Schaus, Taylor, and
  Borisy}}]{schaus_actin_selforg}
\bibinfo{author}{\bibfnamefont{T.~E.} \bibnamefont{Schaus}},
  \bibinfo{author}{\bibfnamefont{E.~W.} \bibnamefont{Taylor}},
  \bibnamefont{and} \bibinfo{author}{\bibfnamefont{G.~G.}
  \bibnamefont{Borisy}}, \bibinfo{journal}{Proc. Natl. Acad. Sci.}
  \textbf{\bibinfo{volume}{104}}, \bibinfo{pages}{7086} (\bibinfo{year}{2007}).

\bibitem[{\citenamefont{Doubrovinski and Kruse}(2007)}]{kruse_actinasters}
\bibinfo{author}{\bibfnamefont{K.}~\bibnamefont{Doubrovinski}}
  \bibnamefont{and} \bibinfo{author}{\bibfnamefont{K.}~\bibnamefont{Kruse}},
  \bibinfo{journal}{Phys. Rev. Lett.} \textbf{\bibinfo{volume}{99}},
  \bibinfo{pages}{228104} (\bibinfo{year}{2007}).

\bibitem[{\citenamefont{Kruse and
  J\"{u}licher}(2003)}]{kruse_juelicher_actin_selforg_1}
\bibinfo{author}{\bibfnamefont{K.}~\bibnamefont{Kruse}} \bibnamefont{and}
  \bibinfo{author}{\bibfnamefont{F.}~\bibnamefont{J\"{u}licher}},
  \bibinfo{journal}{Phys. Rev. E} \textbf{\bibinfo{volume}{67}},
  \bibinfo{pages}{051913} (\bibinfo{year}{2003}).

\bibitem[{\citenamefont{Kruse et~al.}(2004)}]{kruse_juelicher_actin_selforg_2}
\bibinfo{author}{\bibfnamefont{K.}~\bibnamefont{Kruse}} \bibnamefont{et~al.},
  \bibinfo{journal}{Phys. Rev. Lett.} \textbf{\bibinfo{volume}{92}},
  \bibinfo{pages}{078101} (\bibinfo{year}{2004}).

\bibitem[{\citenamefont{Astr\"{o}m et~al.}(2009)\citenamefont{Astr\"{o}m,
  Kumar, and Karttunen}}]{astrom_filaments+motors}
\bibinfo{author}{\bibfnamefont{J.~A.} \bibnamefont{Astr\"{o}m}},
  \bibinfo{author}{\bibfnamefont{P.~B.~S.} \bibnamefont{Kumar}},
  \bibnamefont{and}
  \bibinfo{author}{\bibfnamefont{M.}~\bibnamefont{Karttunen}},
  \bibinfo{journal}{Soft Matter} \textbf{\bibinfo{volume}{5}},
  \bibinfo{pages}{2869} (\bibinfo{year}{2009}).

\bibitem[{\citenamefont{Hawkins et~al.}(2009)\citenamefont{Hawkins,
  B\'{e}nichou, Piel, and Voituriez}}]{hawkins_voituriez_polarization}
\bibinfo{author}{\bibfnamefont{R. J.}~\bibnamefont{Hawkins}},
  \bibinfo{author}{\bibfnamefont{O.}~\bibnamefont{B\'{e}nichou}},
  \bibinfo{author}{\bibfnamefont{M.}~\bibnamefont{Piel}}, \bibnamefont{and}
  \bibinfo{author}{\bibfnamefont{R.}~\bibnamefont{Voituriez}},
  \bibinfo{journal}{Physical Review E} \textbf{\bibinfo{volume}{80}},
  \bibinfo{pages}{040903} (\bibinfo{year}{2009}).

\bibitem[{\citenamefont{Greulich and Santen}(2010)}]{celltransport_per}
\bibinfo{author}{\bibfnamefont{P.}~\bibnamefont{Greulich}} \bibnamefont{and}
  \bibinfo{author}{\bibfnamefont{L.}~\bibnamefont{Santen}},
  \bibinfo{journal}{Eur. Phys. J. E} \textbf{\bibinfo{volume}{32}},
  \bibinfo{pages}{191} (\bibinfo{year}{2010}).

\bibitem[{\citenamefont{Hinow et~al.}(2009)\citenamefont{Hinow, Rezania, and
  Tuszy\'nski}}]{microtubule_mathmodel}
\bibinfo{author}{\bibfnamefont{P.}~\bibnamefont{Hinow}},
  \bibinfo{author}{\bibfnamefont{V.}~\bibnamefont{Rezania}}, \bibnamefont{and}
  \bibinfo{author}{\bibfnamefont{J.~A.} \bibnamefont{Tuszy\'nski}},
  \bibinfo{journal}{Phys. Rev. E} \textbf{\bibinfo{volume}{80}},
  \bibinfo{pages}{31904} (\bibinfo{year}{2009}).

\bibitem[{Not()}]{Note1}
\bibinfo{note}{This normalization is chosen to separate the influence of
  dynamics (in $\protect \mathbf G(\protect \mathbf x-\protect \mathbf x')$ )
  and geometry (in $\rho _n(\protect \mathbf x)$)}.

\bibitem[{\citenamefont{Kruse et~al.}(2005)}]{kruse2005}
\bibinfo{author}{\bibfnamefont{K.}~\bibnamefont{Kruse}} \bibnamefont{et~al.},
  \bibinfo{journal}{EPJ E} \textbf{\bibinfo{volume}{16}}, \bibinfo{pages}{5}
  (\bibinfo{year}{2005}).

\bibitem[{\citenamefont{{Lee} and Kardar}(2001)}]{lee_filament_selforg}
\bibinfo{author}{\bibfnamefont{H. Y.}~\bibnamefont{{Lee}}} \bibnamefont{and}
  \bibinfo{author}{\bibfnamefont{M.}~\bibnamefont{Kardar}},
  \bibinfo{journal}{Physical Review E} \textbf{\bibinfo{volume}{64}},
  \bibinfo{pages}{056113} (\bibinfo{year}{2001}).

\bibitem[{\citenamefont{Terenna et~al.}(2008)}]{terenna}
\bibinfo{author}{\bibfnamefont{C.~R.} \bibnamefont{Terenna}}
  \bibnamefont{et~al.}, \bibinfo{journal}{Current biology : CB}
  \textbf{\bibinfo{volume}{18}}, \bibinfo{pages}{1748} (\bibinfo{year}{2008}).

\end{thebibliography}

\end{document}

\title{\large Supplementary Material}

\bibliographystyle{apsrev}

\maketitle

\section{Filament and particle dynamics}

Tables \ref{particle_dyn_tab} and \ref{dis_netw_dyn_tab} are summarizing the elementary steps of the particle and filament dynamics. Note that we distinguish between microtubule (MT) and actin dynamics in our model in the following way \cite{alberts}: In contrast to MTs, actin filaments can uncap and depolymerize at the minus end to perform \emph{treadmilling}. MTs do not treadmill but, on the other hand, can perform dynamic instability at the plus end. They can turn to a plus-end depolymerizing state, which can be rescued to turn to a polymerizing state.

For all simulations we used a stochastic random sequential update scheme and continuous space coordinates.

%%%%%%%%%%%%%%%%%%%%%%%%%%%%%%%%%%%%%%%%%%%%%%%%%%%%%%%%%%%%%%%%%%%%%%%%%%%%%%%%%%%%
\begin{table*}[h]
\begin{center}
\begin{tabular}{cccc}
Process & Particle state(s) & Description & Parameter name \\ \hline \\ \vspace{2mm}
\emph{Diffusion} & D &
\begin{minipage}[t]{8cm} Detached particles move in a random direction. Step widths are uniformly distributed between $0$ and $2l_D$.  \end{minipage} & $l_D$ \\ \vspace{2mm}
\emph{Step} & A & 
\begin{minipage}[t]{8cm} Attached particles move to adjacent subunit in (+)-direction.  \end{minipage} & $p$ \\ \vspace{2mm}
\emph{Attachment} & D$\to$A & 
\begin{minipage}[t]{8cm} Particles bind to subunits if their distance is less than $d_{b}$, becoming 'attached'. \end{minipage} & $\omega_a$ \\ \vspace{2mm}
\emph{Detachment} & A$\to$D& 
\begin{minipage}[t]{8cm} Particles detach from filament. \end{minipage} & $\omega_d$ \\ \vspace{2mm}
\end{tabular}
\caption{\label{particle_dyn_tab} Particle dynamics. A='attached'; D='detached'.} 
\end{center} 
\end{table*}
%%%%%%%%%%%%%%%%%%%%%%%%%%%%%%%%%%%%%%%%%%%%%%%%%%%%%%%%%%%%%%%%%%%%%%%%%%%%%%%%%%%%

%%%%%%%%%%%%%%%%%%%%%%%%%%%%%%%%%%%%%%%%%%%%%%%%%%%%%%%%%%%%%%%%%%%%%%%%%%%%%%%%%%%%
\begin{table*}[h]
\begin{center}
\begin{tabular}{ccc}
Process & Description & Probability \\ \hline \\
\emph{Nucleation} & 
\begin{minipage}[t]{8cm} New filament created; arbitrary direction. The process needs nucleations seeds (e.g. $\gamma$-tubulin for microtubules) and monomers. $(-)$-Cap inhibits shrinking.  \end{minipage} & $\begin{array}[t]{l} \omega_n\,\rho_{mon} \,\rho_{nuc} \\ \rho_{nuc} = \mbox{nucl. seeds} \\ \rho_{mon} = \mbox{density of free monomers} \\ \end{array}$  \vspace{1mm} \\
\emph{Growth} & 
\begin{minipage}[t]{8cm} New sub-segment added at $(+)$-end. \end{minipage} & $\omega_g \rho_{mon} $ \vspace{1mm} \\
\emph{Shrinking} & 
\begin{minipage}[t]{8cm} Segment removed at $(-)$-end if not $(-)$-capped or at $(+)$-end if in shrink state. \end{minipage} & $\omega_s^{+/-}$ \vspace{1mm} \\
%\emph{Branching} & 
%\begin{minipage}[t]{3cm} New filament is generated at  existing one; angle between parent filament and branch=$70^o$. (only actin)  \end{minipage} & $\begin{array}{l} \omega_b\,\rho_{act} \,\rho_{ARP} \\ \omega_b=0.0035 lu^3 \end{array}$ \vspace{1mm} \\
\emph{Uncapping} & 
\begin{minipage}[t]{8cm} $(-)$-Cap is removed allowing depolymerization hence (only actin). \end{minipage} & $\omega_u$  \vspace{1mm}  \\
\emph{Catastrophe} & 
\begin{minipage}[t]{8cm} $(+)$-end turns into shrink state. (only MT) \end{minipage} & $\omega_{cat}$  \vspace{1mm} \\
\emph{Rescue} & 
\begin{minipage}[t]{8cm} $(+)$-end turns into growth state again. (only MT). \end{minipage} & $\omega_{res}$ \vspace{1mm}  \\
\end{tabular}
\caption{\label{dis_netw_dyn_tab} Filament dynamics.}
\end{center} 
\end{table*}
%%%%%%%%%%%%%%%%%%%%%%%%%%%%%%%%%%%%%%%%%%%%%%%%%%%%%%%%%%%%%%%%%%%%%%%%%%%%%%%%%%%%

\section{Choice of default parameters}

The default parameters of our model were chosen in accordance with available experimental results or, if not directly accessible, according to established models. In Table \ref{default_parameters_microtubule} the parameters for microtubule and vesicle dynamics are given. For dynamics of actin dynamics, we refer to the work \cite{celltransport_per}, Table 4. 

The measured quantities in Table \ref{default_parameters_microtubule} (column 4) are displayed according to their sources referenced in column 3. These measured quantities are transferred to the time and length scales used in the simulations (column 5).\\

\emph{Comments}:
\begin{itemize}
\item[(i)] The maximal tubulin density is reached, if all microtubules are dissolved. Assuming 26 tubulin dimers per $d_s=16\,nm$ microtubule segments \cite{alberts} (16nm are two twists of the microtubule structure), 20 microtubules per $\mu m^2$ \cite{microtubule_areadens} and free tubulin concentration of 5$\mu M$ in healthy axons \cite{microtubule_mathmodel}, one obtains 59$\mu M$.
\item[(ii)] All the minus ends of microtubules are bound to $\gamma-$tubulin which also serves as nucleation seed. Assuming an average microtubule length of $4 \,\mu m$ \cite{microtubule_length} and 20 microtubules/$\mu m^2$ \cite{microtubule_areadens} one obtains 5 $\gamma -$tubulin nucleation seeds per $\mu m^3$.
\item[(iii)] All nucleation seeds are already present in the beginning. Each microtubule is hence initiated by a growth process of a bare $\gamma -$tubulin.
\item[(iv)] In reference \cite{attachmentrate_hill_1} the binding rate for kinesin in dependence of the distance $x$ of the binding filament is given by $3\kappa \exp(-x^2/2\sigma^2)$ with $\kappa=175 s^{-1}$ and $\sigma=6nm$. The average binding rate within a distance $d_b=80nm$ from a filament of length $L$ hence is $\omega_a = 1/(\pi d_b^2 L) \int_0^L\int_0^{d_b} 2\pi r e^{-\frac{r^2}{2\sigma^2}} \, dr \,dz = 5.9 s^{-1}$. The integration space is due to the cylindrical form of the binding region.
\item[(v)] We adjusted the entry and exit rates of particles such that the observed density in the healthy axon was approximately as in the referenced work.
\item[(vi)] The length of the microtubule depletion zone in the experiments \cite{spira_trap} was about $100\, \mu m$. Due to computational constraints we were able to simulate a system of length $L = 200 \, lu \hat = 20\, \mu m$, giving the length of the depletion zone $L/2 = 100\, lu$. Adjusting to the proportions of  the experimental picture, we chose a cylinder radius $20 \,lu$. However, we also tested other system sizes which did not affect the generic structure of the particle traps.
\end{itemize}

%%%%%%%%%%%%%%%%%%%%%%%%%%%%%%%%%%%%%%%%%%%%%%%%%%%%%%%%%%%%%%%%%%%%%%%%%%%%%%%%%%%%
\begin{table*}[h]
\begin{center}
\begin{tabular}{cccc} \hline \\
\vspace{2mm}  Parameter name & Reference & Reference Value & Model parameters \\ \hline \\
\vspace{2mm} \emph{Filament dynamics}: &  &  &   \\  \hline \\
growth rate $\omega_g$ & \cite{microtubule_mathmodel} & $1.5\,\mu m\,\mu M^{-1}min^{-1}$ & $0.015 \,lu^3 \,tu^{-1}$   \\ 
shrink rate $\omega_s$ & \cite{microtubule_mathmodel} & $50 \mu m\, min^{-1}$ & $0.5 \,tu^{-1}$  \\
microtubule density $\rho_{MT}$ & \cite{microtubule_areadens} & $20\, \mu m^{-2}$ & $0.2 \,lu^{-2}$   \\
max. monomer density $\rho_{mon}$ & $\gamma$-tubulin density \cite{microtubule_areadens,alberts,microtubule_mathmodel} & $59 \,\mu M$ ( See comment (i))& $35\, lu^{-3}$ \\
$\gamma -$tubulin density $\rho_{\gamma}$ &  \cite{microtubule_length,microtubule_areadens} & $5 \,\mu m^{-3}$ (See comment (ii))  & $0.005\, lu^{-3}$   \\ 
nucleation rate $\omega_n$ &  See comment (iii) & $\omega_g\rho_\gamma\rho_{tub}$ & see above   \\
%uncap rate $\omega_u$ & See comment (ii) & See comment (ii) & $0.0001 \,tu^{-1}$  \\
catastrophe rate $\omega_c$ & \cite{microtubule_param_3} & $0.014\,s^{-1}$ & $0.0001 \,tu^{-1}$  \\
rescue rate $\omega_r$ & \cite{microtubule_param_3}  & $0.04\, s^{-1}$ & $0.0004 \,tu^{-1}$ \\  \\ \hline \\  
\vspace{2mm} \emph{Particle dynamics}: &  &  &    \\ \hline  \\
particle radius $r_{p}$ & \cite{vesicle_radius} & 50nm (average) & 0.5 $lu$  \\
binding distance $d_b$ & \cite{alberts} & $80\,nm$  (length of kinesin) & 0.8 $lu$  \\
subunit distance $d_s$ & \cite{alberts} & 16nm (1 stepping period of kinesin) & 0.16 $lu$  \\
step rate $p$ & \cite{axon_vesicles}  & velocity = $1.3 \,\mu m\,s^{-1}$ & $0.13 \,tu^{-1}$  \\
attachment $\omega_a$ & \cite{attachmentrate_hill_1} & $5.9\, s^{-1}$ (See comment (iv)) & 0.059 $tu^{-1}$  \\ 
detachment $\omega_d$ & \cite{axon_vesicles} & run length $\approx 7 \,\mu m$ & $0.0018 \,tu^{-1}$  \\ 
diffusive step length $l_D$  & \cite{vesicle_diff} & diff. const. $D=2.5 \times 10^{-10} \, cm^2 s^{-1}$ & $\sqrt{2\,D\, tu}=\sqrt{0.05} lu$  \\ 
particle density $\rho_p^0$ & \cite{microtubule_vesdens} & $7.6\, \mu m^{-3}$ & $0.0076\,lu^{-3}$ \,\, \\
entry rates $\alpha_{1,2}$ & \cite{microtubule_vesdens} & See comment (v) & $\alpha_1 = 0.15\,tu^{-1}, \alpha_2=0.012 \,tu^{-1}$ \\
exit rates $\beta_{1,2}$ & \cite{microtubule_vesdens} & See comment (v) & $\beta_{1} = 0.003\,tu^{-1},  \beta_2 = 0.2\,tu^{-1}$ \\
System size & See comment (vi) & See comment (vi) & length $L= 200 \,lu$, radius $R= 20 \, lu$ \\
MT depletion zone & See comment (vi) & See comment (vi) & $L/2$ 
\end{tabular}
\caption{\label{default_parameters_microtubule} Default parameters of the model which are biologically motivated by transport of vesicles in axons. The referenced values are either based on experimental data or existing models for microtubule dynamics \cite{microtubule_mathmodel}. Model parameters are chosen to be in the order of magnitude of referenced values, fitted to time and space scale of the simulations. Length scale: $1\,lu$ = 100nm = $2r_{p}$   $\Rightarrow 1\mu M=0.6\,lu^{-3}$.
Time scale: $1 \,tu=\Delta t=0.01s$ which is one time step in the simulations. We consider a cylindrically shaped system length $L=200\,lu$ and a radius of $L/10$. In the transected state, it has a spherically shaped tip and a neck region right of the middle with a minimum radius $0.6 L$.}
\end{center} 
\end{table*}
%%%%%%%%%%%%%%%%%%%%%%%%%%%%%%%%%%%%%%%%%%%%%%%%%%%%%%%%%%%%%%%%%%%%%%%%%%%%%%%%%%%%

%%%%%%%%%%%%%%%%%%%%%%%%%%%%%%%%%%%%%%
\section{Time evolution of the simulations: transected axon}

Here we show some intermediate time configurations of particles and filaments both in the case of a plain cylindrical geometry (with a round tip) and a cylinder with a tapered region, mimicking the geometry of the axon more faithfully. One observes that in both cases accumulations of plus particles occur left of the depletion area and at the tip, similar to the experiments. However for the plain cylindrical geometry it takes longer to achieve this state.

One also observes that in the beginning, the parallel bundles are also present in the right part of the system, while accumulation of vesicles already begins. Hence the accumulation is \emph{not} due to a diffusive bottleneck in the middle of the system. 

%%%%%%%%%%%%%%%%%%%%%%%%%%%%%%%%%%%%%%%%%%%%%%
\begin{figure*}[h]
\includegraphics[width=0.9\columnwidth]{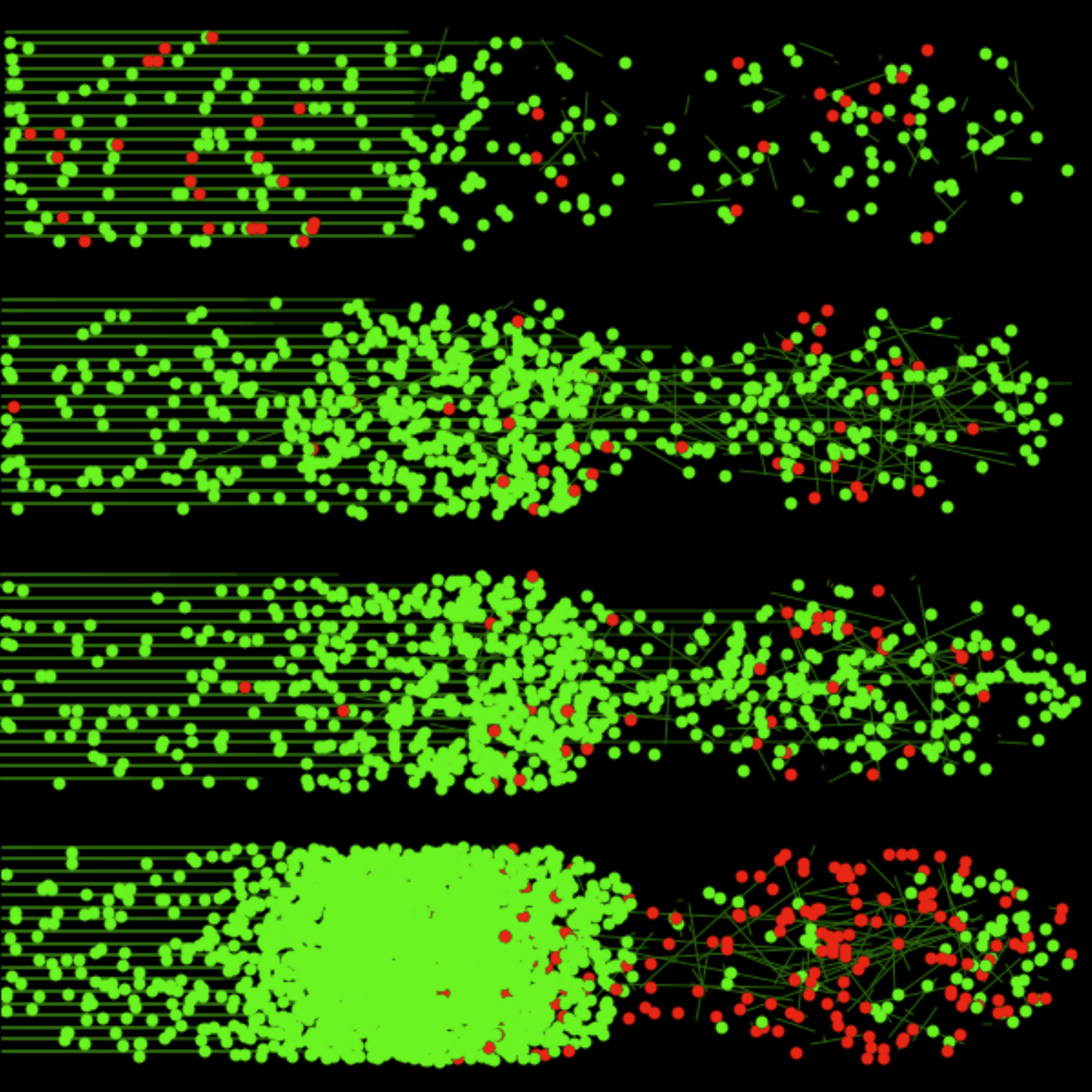}
\includegraphics[width=0.9\columnwidth]{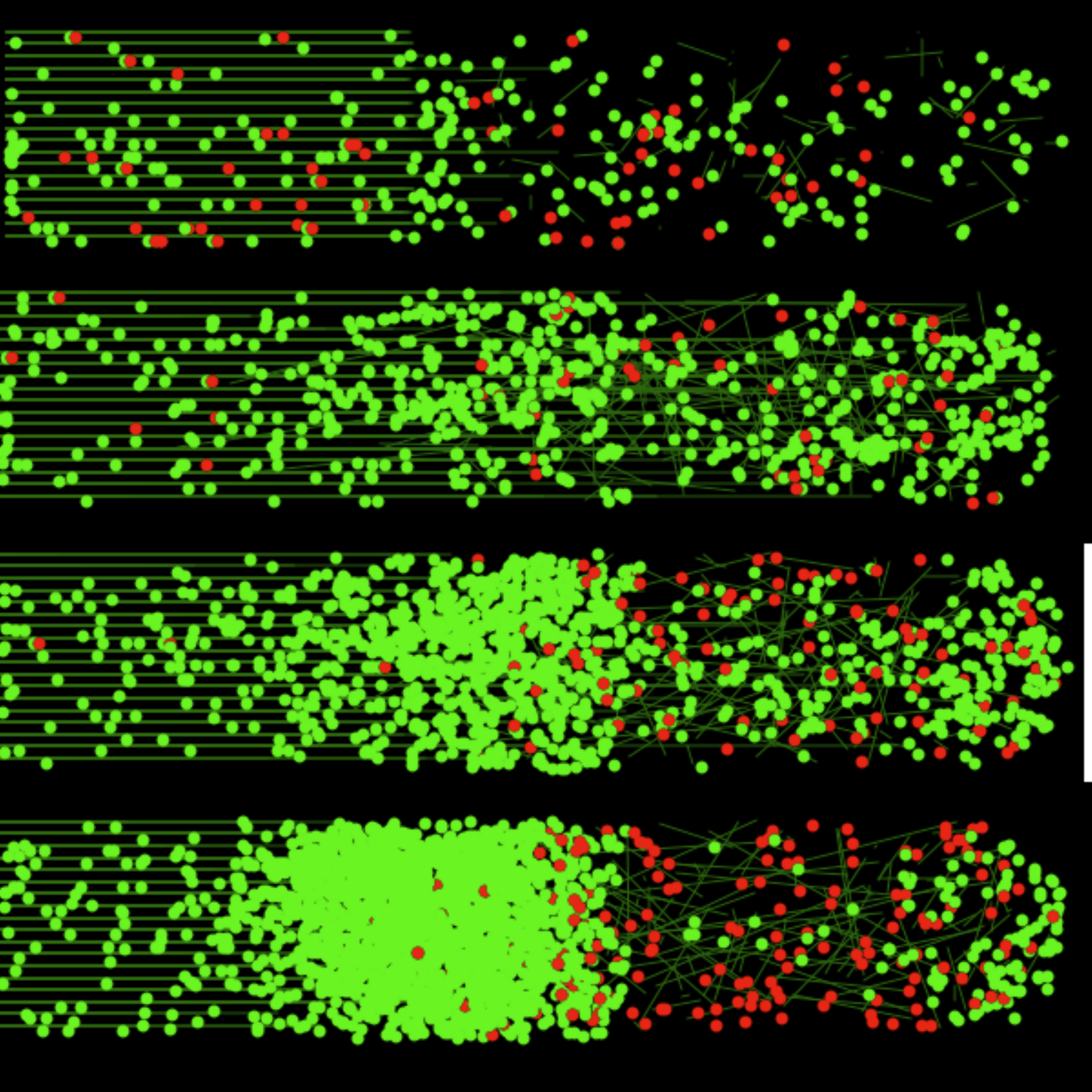}
\caption{Configurations of particles and microtubules for simulations of the model at distinct times. (a) Geometry of a transected axon. (a.1) t=200 (a.2) t=3000 (a.3) t=5000 (a.4) t=24000. (b) Cylindrical geometry. (b.1) t=200 (b.2) t=5000 (b.3) t=20000 (b.4) t=32000. The qualitative structure does not depend on the details of the geometry.}
\end{figure*}

\newpage

\bibliography{library}